\documentclass[final,1p,times]{elsarticle} 

\usepackage{graphicx}
\usepackage{amssymb} 
\usepackage{amsthm} 
\usepackage{lineno}

\usepackage{hyperref}

\usepackage{sidecap}

\newcommand{\Dzero}{{\rm D^0}}
\newcommand{\Dstar}{{\rm D^{*+}}}
\newcommand{\Dplus}{{\rm D^+}}
\newcommand{\sqrtsNN}{\sqrt{s_{\rm \scriptscriptstyle NN}}}
\newcommand{\DtoKpi}{{\rm D^0\to K^-\pi^+}}
\newcommand{\DtoKpipi}{{\rm D^+\to K^-\pi^+\pi^+}}
\newcommand{\DstartoDpi}{{\rm D^{*+}\to D^0\pi^+}}
\journal{Nuclear Physics A} 
\begin{document}
\begin{frontmatter} 
\title{Measurement of the D meson elliptic flow in Pb--Pb collisions\\
  at $\sqrt{s_{\rm NN}} = 2.76~\rm{TeV}$ with ALICE}|

\author{Davide Caffarri (for the ALICE\fnref{col1} Collaboration)}
\fntext[col1] {A list of members of the ALICE Collaboration and acknowledgements can be found at the end of this issue.}
\address{Universit\`a degli Studi di Padova and INFN, Italy}
\vspace{-28pt}
\begin{abstract} 
  We present the measurements of $\Dzero$, $\Dplus$, $\Dstar$ meson
  $v_2$ as well as $\Dzero$ $R_{\rm AA}$ in different directions with
  respect to the estimated reaction plane in Pb--Pb collisions at
  $\sqrtsNN =~2.76~\rm{TeV}$ with the ALICE detector at the LHC.
\end{abstract} 

\end{frontmatter} 

\vspace{-16pt}
\section{Introduction}
On the basis of thermodynamical considerations and Quantum
Chromo-Dynamics (QCD) calculations, nuclear matter, in conditions of
high temperature and density, is expected to undergo a phase transition
to a deconfined state: the Quark
Gluon Plasma (QGP)~\cite{Karsch:2006xs}. These conditions can be
recreated in high energy nucleus--nucleus collisions. 

Non-central nucleus--nucleus collisions are characterized by an initial
geometrical anisotropy with respect to the reaction plane (the plane
defined by the beam direction and the impact parameter). 
As a consequence of the different pressure gradients in the in-plane
and out-of-plane regions, this spatial anisotropy is converted to a momentum anisotropy of produced low-$p_{\rm T}$ particles.
The momentum anisotropy can be quantified
via a Fourier expansion of the azimuthal angle with respect to the
estimated reaction plane. The second coefficient of this expansion $v_2$ is called
elliptic flow~\cite{charm_flow1}. 
The azimuthal anisotropy of heavy-flavour hadron production is sensitive
to the degree of thermalization of heavy quarks in the expanding medium and
to the path length dependence of their energy loss~\cite{charm_flow2,
  charm_flow3, Alberico:2011zy, Gossiaux:2009mk, Uphoff:2011aa,
  He:2012df}.
\vspace{-12pt}
\section{Data sample and event plane determination}
The D meson azimuthal anisotropy measurements
presented in this report are performed with the ALICE
detector~\cite{JINST_ALICE} using the 2011 data
sample collected with a minimum bias trigger given by the coincidence of
signals in the VZERO scintillators and the Silicon Pixel Detector (SPD) and a trigger
tuned to enhance the sample of 50\% most central collisions, based on
the VZERO signal amplitude, which is used to classify the events
according to centrality. 
The analyses have been peroformed with $~9.5 \times 10^{6}$ Pb--Pb
collisions in the centrality class 30--50\%,$~7.1 \times 10^{6}$ in 15--30\%
and $~16 \times 10^{6}$ in 0--7.5\%.

The azimuthal direction of the event plane, which estimates the
reaction plane direction, is determinated from the distribution of charged tracks in
the $ 0 < \eta <0.8$ pseudo-rapidity interval of the Time
Projection Chamber (TPC) using the equation 
\begin{equation}
\Psi = \frac{1}{2} \tan^{-1} \left( \frac{\sum^{N}_{i=0} w_{i} \sin
    2\varphi_{i}}{\sum^{N}_{i=0} w_{i} \cos 2\varphi_{i}}  \right),
\end{equation}
where $\Psi$ is the second harmonic event plane and $\varphi_{i}$ is the
angle of the $i^{th}$ track in the ALICE reference frame. 
The event plane resolution ($R_{2}$) was computed considering each event as
splitted in 2 sub-events, made of randomly combined tracks and
associated to the correspondent event plane angle, following the
prescription in~\cite{epSubEv}. The resulting event plane is flat to about 2\%
level and its resolution is 0.86 in the 30--50\% centrality class, 0.9 in
15--30\% and 0.75 in 0--7.5\%.
\vspace{-12pt}
\section{D meson reconstruction, $v_2$ and $R_{\rm AA}$ determination}
D meson reconstruction in ALICE is based on the invariant mass
analysis of fully reconstructed hadronic decay topologies of these
channels: $\DtoKpi$, $\DtoKpipi$ and $\DstartoDpi$. The
separation of about a few hundred $\mu$m between the primary and the secondary
vertices, peculiar of $\Dzero$ and $\Dplus$ decay, is exploited to reduce the
combinatorial background. In ALICE, charged tracks are reconstructed
with a six layers silicon tracker (Inner Tracking System) and the Time
Projection Chamber (TPC).
In particular, the Silicon Pixel Detector (SPD) provides a measurement of track impact
parameter to the primary vertex with a resolution better than 65
$\mu$m for tracks with $p_{\rm T} > 1~{\rm GeV}/c$. Particle
identification is provided by the time of flight measurement in the
Time-Of-Flight detector for kaons with $p_{T} < 2~{\rm GeV}/c$ and by the
specific energy deposit in the TPC. For high-$p_{\rm T}$ particles, a
selection is applied in order to reject protons without loosing
possible kaons. The contribution of signal candidates with wrong mass assignement to the final state hadrons was found to be
negligible and it does not bias the signal extraction. Further details on
the analysis can be found in~\cite{ALICE:2012ab}. The invariant mass
distribution of reconstructed D meson candidates is considered in the in-plane and
out-of-plane regions defined as $\left[ 0 < \Delta\varphi <
\frac{\pi}{4} \right) \cup \left[ \frac{3\pi}{4} < \Delta\varphi <\pi \right)$
and $\left[ \frac{\pi}{4} < \Delta\varphi <
\frac{3\pi}{4} \right)$ respectively, where $\Delta\varphi$ is the
azimuthal angle of the reconstructed candidate with respect to the event
plane $\Psi$. An invariant mass analysis, based on a Gaussian fit, is used to
obtain the signal yield in the two regions ($N_{\rm IN}$, $N_{\rm
  OUT}$). Assuming that the reconstruction and selection efficiency is
indipendent of $\Delta\varphi$, it is possible to
directly compute the elliptic flow from the formula (\ref{eq_v2_Raa},
left), where $R_{2}$ is the event plane resolution.
\begin{equation}
\label{eq_v2_Raa}
v_{2} = \frac{\pi}{4} \frac{1}{R_{2}} \frac{N_{\rm IN} - N_{\rm OUT}}{N_{\rm IN} +
  N_{\rm OUT}}, ~~~ R_{\rm AA} (p_{\rm T}) = \frac { {\rm d}N_{AA}/ {\rm d} p_{\rm T}}
{\langle T_{\rm{AA}} \rangle \times {\rm d}\sigma_{pp}/{\rm d} p_{\rm T}}
\end{equation} 
The $R_{ \rm AA}$ measurement is based on the comparison between the
yield measured in AA collisions and the cross section measured in pp
collisions, scaled by the overlap nuclear function, as shown in
formula (\ref{eq_v2_Raa}, right). For this analysis, the yields $N_{\rm
  IN}$ and $N_{\rm OUT}$ were corrected for acceptance and efficiency as a function of $p_{\rm T}$,
using  PYTHIA + HIJING Monte Carlo simulations~\cite{ALICE:2011aa}.
Details on the correction for the secondary D meson and the pp measurement, used as
reference, can be found in~\cite{ALICE:2011aa}.
\begin{figure}[t]
\begin{center}
\includegraphics[width=0.48\textwidth]{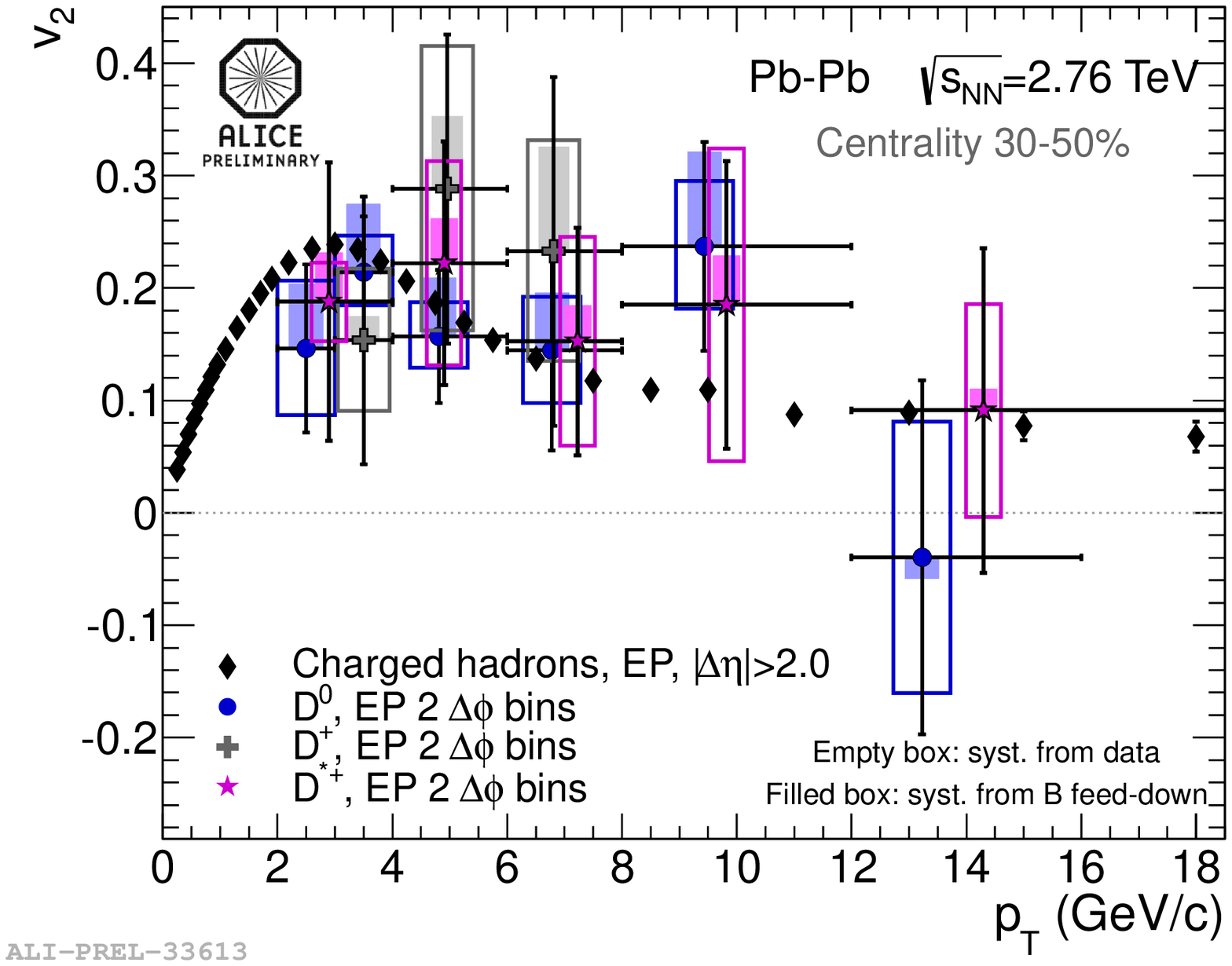}
\includegraphics[width=0.48\textwidth]{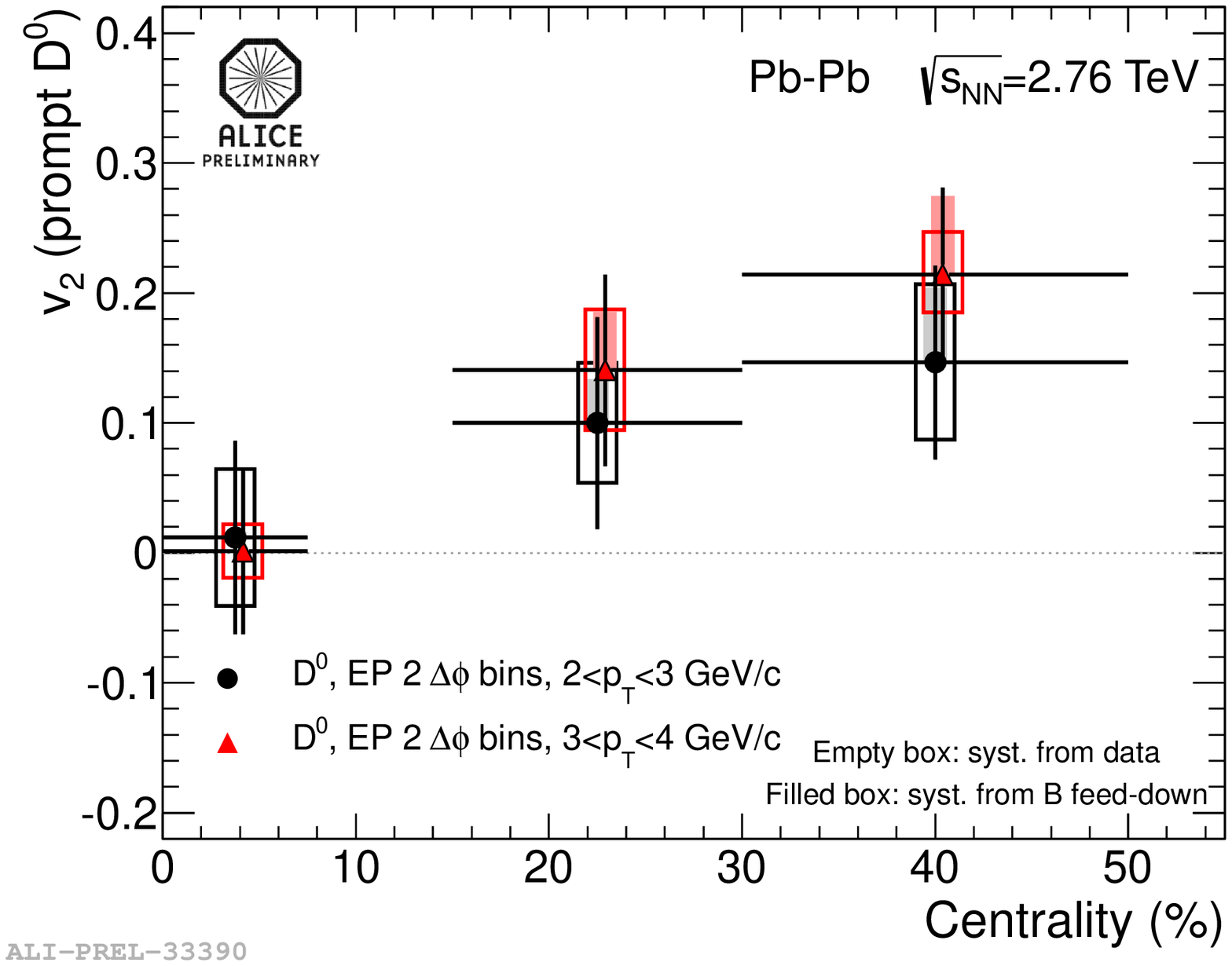}
\end{center}
\caption{ (left) $\Dzero$, $\Dplus$ and $\Dstar$ $v_2$ measured by ALICE in
the 30-50\% centrality class. Empty boxes show systematic uncertaintites
from data and the shaded area the B feed-one one. Black points shows the
charged hadron $v_2$. (right) $\Dzero$ $v_2$ as a function of centrality for
the bins $2 < p_{\rm T} < 3~{\rm GeV}/c$, $3< p_{\rm T} < 4~{\rm
  GeV}/c$. Empty boxes show systematic uncertaintites
from data and the shaded areas the B feed-down one.}
\label{fig:DmesV2}
\end{figure}

Several sources of systematic uncertainties were considered in 
both analyses. The main contributions come from uncertainties on the
yield extraction and from the topological cut selection. The former is
estimated using different background functions in the fitting
precedure, different mass range, and a bin counting method. 
The analysis was also repeated with three different sets of topological
cuts. Yield extraction and cut variation systematic uncertainties are
estimated to be at maximum 0.05 absolute value, for the $v_2$ analysis.
The signal sample considered contains a fraction of D mesons coming from
B decays, thus the measured elliptic flow is a combination of prompt
and secondary D meson anisotropy. The feed-down correction uses as
input pQCD calculations of B production, as explained in~\cite{ALICE:2012ab}. 
The resulting systematic uncertainty on $v_2$ is up to $+23\%$,
including a conservative variation of the unknown $R_{\rm AA}$ and
$v_2$ of D meson from B decays. 
The centrality dependence of the event plane resolution in the
centrality range considered was also taken into account, together
with the uncertainty on the resolution. For the 30-50\% centrality class, these
uncertainties were found to be $\pm 3\%$ and  $+7\%$ respectively.
The small difference of efficiencies in the two $\Delta\varphi$ region was
found to give a negligible contribution to the $v_2$ uncertainties.  
\vspace{-12pt}
\section{Results}
The D meson anisotropy was measured in the 30--50\% centrality
class: for $\Dzero$ in the range $2<
p_{\rm T} < 16~ {\rm GeV}/c$, $\Dplus$ in $3< p_{\rm T} < 8~ {\rm GeV}/c$  and $\Dstar$ in $2<
p_{\rm T} < 20~ {\rm GeV}/c$. The three results are compatible within
statistical uncertainties and all are compatible with the ALICE
measurement of charged hadrons in the same rapidity region
(Fig.~\ref{fig:DmesV2}, left). 
The $\Dzero$ $v_2$ measurement was also performed in the centrality
classes 15--30\% and 0--7.5\% in the range $2<
p_{\rm T} < 16~{\rm GeV}/c$. The results for the first two $p_{\rm T}$
intervals show a hint of increasing $v_2$ from central to
semi-peripheral collisions, as reported in Fig.\ref{fig:DmesV2},
right.
$v_2$ was measured using two-particles correlation methods as
well. The results are consistent with those based on the event plane.
The measurement of the D meson anisotropy was compared to
theoretical predictions~\cite{ZaidaQMproc}: it was observed that a simultaneous description of
the anisotropy and of the $R_{\rm AA}$ suppression is challenging for models.

\begin{figure}[t]
\begin{center}
\includegraphics[width=0.78\textwidth]{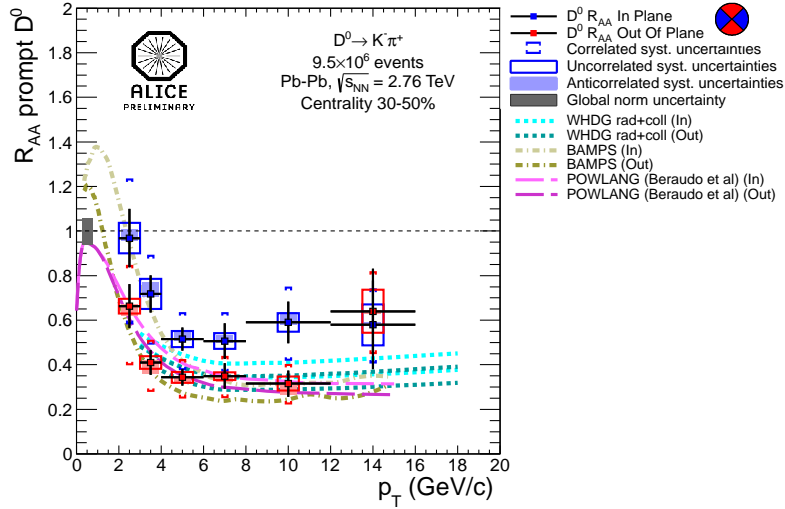}
\end{center}
\caption{$\Dzero$ $R_{\rm AA}$ vs event plane for the 30--50\% centrality
  class. Empty boxes show the uncorrelated systematic uncertainties
  between the two measurements, the empty brackets the correlated systematic
  uncertainties that would shift both measurements in the same
  direction and shaded areas the anticorrelated uncertainties that
  would shift the measurements in opposite directions.}
\label{fig:raavsEP}
\end{figure}

The $\Dzero$ $R_{\rm AA}$ in the two azimuthal regions, in-plane and
out-of-plane, was measured in the range $2< p_{\rm T} < 16~{\rm
  GeV}/c$. The measurement shows more suppression in
the out-of-plane region with respect to the in-plane one, suggesting a
path length dependence of heavy quark energy loss, which is expected
to be the dominant effect at high $p_{\rm T}$
(Fig.~\ref{fig:raavsEP}). At low $p_{\rm T}$, elliptic flow may also
contribute to the observed anisotropy.
In Fig.~\ref{fig:raavsEP} we compare the $\Dzero$ $R_{\rm AA}$ in the
two azimuthal regions with theoretical predictions. The WHDG~\cite{Horowitz:2011cv} and
POWLANG~\cite{Alberico:2011zy} models show a good agreement
with out-of-plane results while they somewhat oversuppress the
in-plane yields. This high-$p_{\rm T}$ feature is present also for the
BAMPS model~\cite{Uphoff:2011aa}, which instead can describe the data
in the low-$p_{\rm T}$ range.
\vspace{-12pt}
\section{Conclusions}
The measurement of D meson $v_2$ in Pb--Pb collisions at $\sqrt{s_{\rm
    NN}} = 2.76~{\rm TeV}$ with ALICE was presented. The results
indicate $v_2 > 0$ in the $p_{\rm T}$ range $2 < p_{\rm T} < 6~{\rm GeV}/c$ with a 3$\sigma$ significance in the 30-50\% centrality
class.  A hint of centrality dependence is observed.
The measurement of the $\Dzero$ $R_{\rm AA}$ around two orthogonal
directions with respect to the event plane, in the 30--50\%
centrality class, indicates a larger suppression in the out-of-plane
direction, where the initial overlap region of the two nuclei is more
elonged, suggesting a path lenght dependence of heavy-quark energy
loss. 

\end{document}